# Nearly quantized conductance plateau of vortex zero mode in an iron-based superconductor

Shiyu Zhu[1,2†], Lingyuan Kong[1,2†], Lu Cao[1,2†], Hui Chen[1,2†], Michał Papaj[5], Shixuan Du[1,2,3,6], Yuqing Xing[1,2], Wenyao Liu[1,2], Dongfei Wang[1,2], Chengmin Shen[1,3], Fazhi Yang[1,2], John Schneeloch[4], Ruidan Zhong[4], Genda Gu[4], Liang Fu[5], Yu-Yang Zhang[2,1,3*], Hong Ding[1,2,3,6*], and Hong-Jun Gao[1,2,3,6*]

[1]Beijing National Laboratory for Condensed Matter Physics and Institute of Physics, Chinese Academy of Sciences, Beijing 100190, China

[2]School of Physical Sciences, University of Chinese Academy of Sciences, Beijing 100190, China

[3]CAS Center for Excellence in Topological Quantum Computation, University of Chinese Academy of Sciences, Beijing 100190, China

[4]Condensed Matter Physics and Materials Science Department, Brookhaven National Laboratory, Upton, New York 11973, USA

[5]Department of Physics, Massachusetts Institute of Technology, Cambridge, Massachusetts 02139, USA

[6]Songshan Lake Materials Laboratory, Dongguan, Guangdong 523808, China

†These authors contributed equally to this work

*Correspondence to: hjgao@iphy.ac.cn, dingh@iphy.ac.cn, zhangyuyang@ucas.ac.cn

**Majorana zero-modes (MZMs) are spatially-localized zero-energy fractional quasiparticles with non-Abelian braiding statistics that hold a great promise for topological quantum computing. Due to its particle-antiparticle equivalence, an MZM exhibits robust resonant Andreev reflection and $2e^2/h$ quantized conductance at low temperature. By utilizing variable-tunnel-coupled scanning tunneling spectroscopy, we study tunneling conductance of vortex bound states on FeTe$_{0.55}$Se$_{0.45}$ superconductors. We report observations of conductance plateaus as a function of tunnel coupling for zero-energy vortex bound states with values close to or even reaching the $2e^2/h$ quantum conductance. In contrast, no such plateau behaviors were observed on either finite energy Caroli-de Genne-Matricon bound states or in the continuum of electronic states outside the superconducting gap. This unique behavior of the zero-mode conductance reaching a plateau strongly supports the existence of MZMs in this iron-based superconductor, which serves as a promising single-material platform for Majorana braiding at a relatively high temperature.**

Majorana zero-modes (MZMs) obey non-Abelian statistics and have potential applications in topological quantum computation (*1, 2*). In the past two decades, MZMs have been predicted in p-wave superconductors (*3, 4*) and spin-orbit-coupled materials proximitized by s-wave superconductors (*5-8*). Experimental evidence for MZMs has been observed in various systems, ranging from semiconductor-superconductor nanowires (*9, 10*), topological insulator-superconductor heterostructures (*11*), atomic chains on superconducting substrate (*12, 13*). Recently, fully gapped bulk iron-based superconductors have emerged as a new single-material platform for MZMs (*14, 15*). Evidence for MZMs in topological vortices on the surface of FeTe$_{0.55}$Se$_{0.45}$ has been found by scanning tunneling microscopy/spectroscopy (STM/S) (*16-18*).

A hallmark of an MZM is its $2e^2/h$-quantized conductance plateau (*19, 20*) independent of the strength of tunnel coupling at sufficiently low temperature. This quantized Majorana conductance results from perfect resonant Andreev reflection guaranteed by the inherent particle-hole symmetric nature of MZM (*2*). A quantized conductance plateau has been observed in an InSb-Al nanowire system, consistent with the existence of MZMs (*21*). However, the solo realization in the complicated nanowire system is not enough to substantiate MZMs. Particularly, some alternative explanations have not been ruled out, *e.g.* the partially separated Andreev bound state (ps-ABS) can also lead to a quantized conductance plateau which is topological trivial (*22-24*). Recently, STM/S experiments observed zero-bias conductance peaks (ZBCPs) in a simple system of iron-based superconductors (*16-18, 25*). Its large topological gap (*e.g.* $\delta_{Fe(Te,Se)} \sim \Delta^2/E_F \sim 0.7$ meV) offer a clean way for observing Majorana quantized conductance without contamination from other low-lying Caroli-de Gennes-Matricon bound states (CBSs) (*16, 18*).

Motivated by the above prospects, we employ a variable tunnel coupling STM/S method to study the Majorana conductance over a large range of tip-sample distance in vortex cores of FeTe$_{0.55}$Se$_{0.45}$ (Fig. 1A). The effective electron temperature of our STM is 377 mK as calibrated by tunneling into aluminum (Fig. S1 of (*26*)). In STM/S, the tunnel coupling can be continuously tuned by changing the tip-sample distance (*d*), which correlates with the tunnel-barrier conductance ($G_N \equiv I_t/V_s$, $I_t$ is the tunneling current. $V_s$ is the setpoint voltage) (*16*). With a 2 T magnetic field applied perpendicular to the sample surface, a sharp ZBCP is observed at a vortex core (Fig. 1B). This ZBCP neither disperses nor splits across the vortex core, as expected for an isolated MZM in a quantum-limited vortex (*16-18, 25*). We perform tunnel-coupling dependent measurement on the observed ZBCP. By putting the STM tip at the center of a topological vortex (*18*), we record a set of $dI/dV$ spectra with different tip-sample distances (Fig. 1C). The ZBCP remains a well-defined peak located at the zero energy [voltage offset calibration under different tunnel couplings is discussed in (*26*)]. More significantly, we observe a unique behavior of ZBCP under different $G_N$ (Fig. 1C): the ZBCP peak height saturates at a relatively high tunnel coupling, while the high-bias conductance outside the superconducting gap increases monotonically as a function of $G_N$. This behavior can be better visualized in a three-dimensional plot (Fig. 1D) and a color-scale plot (Fig. 1E) that expands the tunnel coupling dependent spectra on an extra axis of $G_N$. It is clear that the zero-bias conductance reaches a plateau when $G_N$ is around 0.3 $G_0$ ($G_0 \equiv 2e^2/h$). Two plots of conductance curves as a function of $G_N$ are extracted from Figs. 1C-E. The zero-bias conductance barely changes over a wide range of $G_N$ (0.3 $G_0 \sim 0.9$ $G_0$); the average plateau conductance ($G_P$) is equal to 0.64 $G_0$ (Fig. 1F). In contrast, the high-bias conductance at ±1.5 meV and -4.5 meV (Fig. 1G) changes by a factor of more than three as the tip-sample distance varies.

In order to examine the particle-hole-symmetric nature of the MZMs, we compare and contrast conductance behavior of zero-energy MZMs and finite-energy Caroli-de Gennes-Matricon bound states (CBSs). As demonstrated previously, there are two distinct types of vortices, *i.e.* topological (ordinary) vortices with (without) MZM, differing by a half-integer level shift of vortex bound states (*18*). Firstly, we perform a tunnel-coupling dependent measurement on a topological vortex (Figs. 2A-B), which shows an MZM and the first CBS level located at 0 meV and ± 0.31 meV, respectively. In contrast to MZM, we find that the conductance of finite-energy CBS keeps increasing with $G_N$ and shows no plateau. We also carry out measurements on an ordinary vortex. A d$I$/d$V$ spectrum shows a CBS with half-odd integer quantization (Fig. 2C), in which the first three levels of CBSs located at ±0.13 meV, ±0.39 meV and ±0.65 meV, respectively. Again, the conductance values of all the CBSs keep increasing and have no plateau feature in the tunnel-coupling dependent measurement (Fig. 2D). As another check, we repeat the measurement for the same location at zero magnetic field (Figs. 2E-F), and observe a hard superconducting gap. The zero-bias conductance and the high-bias conductance are plotted as functions of $G_N$ (middle and bottom panels of Fig. 2F, respectively). It is evident that both curves keep increasing as the tunnel coupling increases. This observation can be confirmed in a z-offset plot, with a larger z-offset corresponding to a smaller tip-sample distance [Fig. S3 of (*26*)]. Therefore, the conductance plateau feature has only been observed in ZBCP, which indicates the unique behavior of Majorana modes.

The plateau behavior of the zero-bias conductance provides evidence for the Majorana-induced resonant Andreev reflection (*19, 20*). It has been well understood that a perfect transmission of electrons can occur in a symmetric double-barrier system via resonant tunneling through a single quasistationary bound state (Fig. 2G). The transmission on resonance is $e^2/h$ independent of tunnel coupling, as long as it is identical for the two barriers (*27, 28*). In the case of electron tunneling from a normal electrode through a barrier into a superconductor, the Andreev reflection process (*29*) converts the incident electron into an outgoing hole in the same electrode, resulting in a double-barrier system in the particle-hole Hilbert space. Moreover, in the case of Andreev reflection via a single MZM, the equal amplitude of particle/hole components, due to the particle-antiparticle equivalence of MZM, ensures an identical tunnel coupling with electron and hole in the same electrode ($\Gamma_e = \Gamma_h$) (Fig. 2H). Thus, the resonant Andreev reflection mediated by an MZM leads to a $2e^2/h$-quantized zero-bias conductance plateau, independent of the strength of tunnel coupling at the zero temperature (*19, 20, 30, 31*). However, low-energy CBSs (*32, 33*) and other trivial sub-gap states (*24*) do not have the Majorana symmetry, resulting in unequal weights for electron/hole components. The relationship of $\Gamma_e = \Gamma_h$ is broken in a CBS-mediated Andreev reflection (Fig. 2I), which yields the absence of a conductance plateau (middle panel in Fig. 2B). Moreover, the observed zero-bias conductance plateau in the vortex core disappears after the magnetic field is removed (Figs. 3E-F), hence cannot be attributed to quantum ballistic transport (*34-39*).

The plateau behavior of ZBCPs has been observed repeatedly in many topological vortices [31 plateau features out of 60 measurements (*40*)]. We perform a statistical analysis of the observed plateau values $G_p$ and find that most values of $G_p$ are located around 40% - 60% of $G_0 = 2e^2/h$ (Fig. 3A). In one case, the plateau conductance reaches $G_0$ (Figs. 3B-D). Both instrumental broadening and quasiparticle poisoning in our system can potentially induce deviation of $G_p$ from the theoretical quantized value $2e^2/h$. To examine the possible effect of instrumental broadening on Majorana conductance plateau, we deliberately increase the instrumental broadening by varying the modulation voltage ($V_{mod}$), which defined by the zero-to-peak amplitude of lock-in excitation. This allows us to study the $V_{mod}$-evolution of the Majorana conductance plateau on a given topological vortex (Figs. 3E-F). We find clearly that larger $V_{mod}$ leads to stronger suppression of $G_P$ of MZM. In addition, we notice that the values of conductance plateaus at different vortices can also have some variations. It is found the $G_P$ are correlated with the full width of half maximum (FWHM) of the ZBCP measured under a large tip-sample distance limit. We find that $G_P$ decreases with increasing FWHM (Fig. 3G) [detailed data are shown in Fig. S5]. It indicates that the quasiparticle poisoning effect (*41, 42*) might also play a role on reducing the conductance plateau value, as the poisoning rate is expected to be spatially non-uniform in FeTe$_{0.55}$Se$_{0.45}$ with intrinsic inhomogeneities.

We also check the reversibility of the process of varying tunneling coupling in STM, and find that both the topography and the conductance plateau can be reproduced during two repeated sequences of varying tunneling coupling [Fig. S8 of (*26*)], indicating the absence of irreversible damage of the tip and the sample during measurements. We noted that there are other mechanisms related to zero-bias conductance plateau, *e.g.* inhomogeneity induced ps-ABS (*22-24*) and class-D weak antilocalization (*43*), further understanding of our experiments needs more theoretical efforts. Our observation of a zero-bias conductance plateau in the two-dimensional vortex case, which approaches to the quantized conductance of $2e^2/h$ in the cleanest topological vortex in FeTe$_{0.55}$Se$_{0.45}$, provides the first spatially-resolved spectroscopic evidence for Majorana-induced resonant electron transmission into a bulk superconductor, moving one step further towards the braiding operation applicable to topological quantum computation.

Note added: We noticed that about two months after the submission of this work, a related result in (Li$_{0.84}$Fe$_{0.16}$)OHFeSe was submitted and published (*44*).

ACKNOWLEDGEMENTS

We thank Patrick A. Lee, Chun-Xiao Liu, Gang Su for helpful discussions, and Ai-Wei Wang, Jia-Hao Yan and Qing Huan for technical assistance. The work at IOP is supported by grants from the Ministry of Science and Technology of China (2015CB921000, 2015CB921300, 2016YFA0202300), the National Natural Science Foundation of China (11234014, 61888102), and the Chinese Academy of Sciences (XDB28000000, XDB07000000, 112111KYSB20160061). L.F. and G.G are supported by US DOE (DE-SC0010526, DE-SC0012704, respectively). J.S. and R.Z. are supported by the Center for Emergent Superconductivity, an EFRC funded by the US DOE.

**Author Contributions:** H.-J. G. and H.D. designed STM experiments. S.Z., L.C., H.C., Y.X. and Y.Z. performed STM experiments with assistance of W.L., F.Y. and C. S. J.S., R.Z., and G.G. provided samples. L.F. provided theoretical explanations. Y.Z., S.D., S.Z. and L.K. processed experimental data and wrote the manuscript. All the authors participated in analyzing experimental data, plotting figures, and writing the manuscript. H.D. and H.-J. G. supervised the project.

**Competing interests:** The authors declare that they have no competing interests.

**Data and materials availability:** The data presented in this paper can be found in the supplementary materials.

SUPPLEMENTARY MATERIALS
Materials and Methods
Supplementary Text
Figs. S1 to S9
References (*45 - 47*)

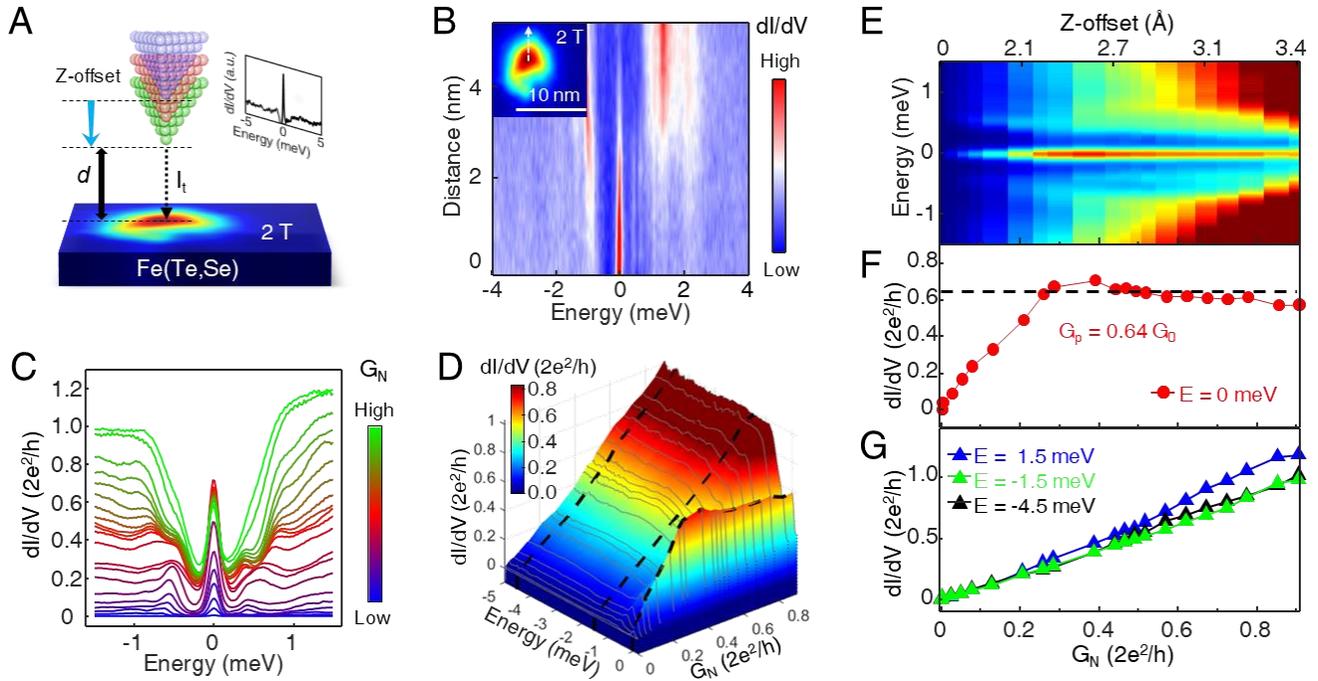

**Fig. 1. Zero-bias conductance plateau observed on FeTe$_{0.55}$Se$_{0.45}$.** (**A**) A schematic of variable tunnel coupling STM/S method. A zero-bias conductance map under 2.0 T is shown on a sample surface. A *dI/dV* spectrum measured at the center of the vortex core ($V_s$ = -5 mV, $I_t$ = 500 pA, $V_{mod}$ = 0.02 mV) is shown in the right-top panel, in which a sharp zero-bias conductance peak (ZBCP) is observed. When the tunneling current ($I_t$) is adjusted by the STM regulation loop, the tunnel coupling between the STM tip and the MZM can be tuned continuously by the tip-sample distance (*d*). Larger tunnel coupling corresponds to smaller *d* and larger tunneling-barrier conductance ($G_N = I_t/V_s$, $V_s$ is the setpoint voltage). Z-offset can be read out simultaneously, which indicates the absolute z-direction motion of the STM tip. (**B**) A line-cut intensity plot along the dashed white arrow in the inset, measured from the same vortex shown in (A), showing a stable MZM across the vortex core. (**C**) An overlapping plot of *dI/dV* spectra under different tunnel coupling values parameterized in $G_N$. The blue curve is measured under the smallest $G_N$ while the green curve with the largest $G_N$. (**D**) A three-dimensional plot of tunnel coupling dependent measurement, *dI/dV(E, $G_N$)*. For clarity, only the data points in the energy range of [-5.0, 0.2] meV are shown. (**E**) A color-scale plot of (C) within the energy range of [-1.5, 1.5] meV that expands the spectra as a function of $G_N$. The z-offset information, which was taken simultaneously by STM, is also labeled at the upper axis. The maximum distance the tip approached is 3.4 Å. This plot shares the same color-bar with (D). (**F**) A horizontal line-cut at the zero-bias from (E). The conductance curve shows a plateau behavior with its plateau conductance ($G_P$) equals to (0.64 ± 0.04) $G_0$. (**G**) Horizontal line-cuts at high-bias values from (E). The absence of a conductance plateau on these curves indicates the conventional tunneling behavior at the energy of continues states.

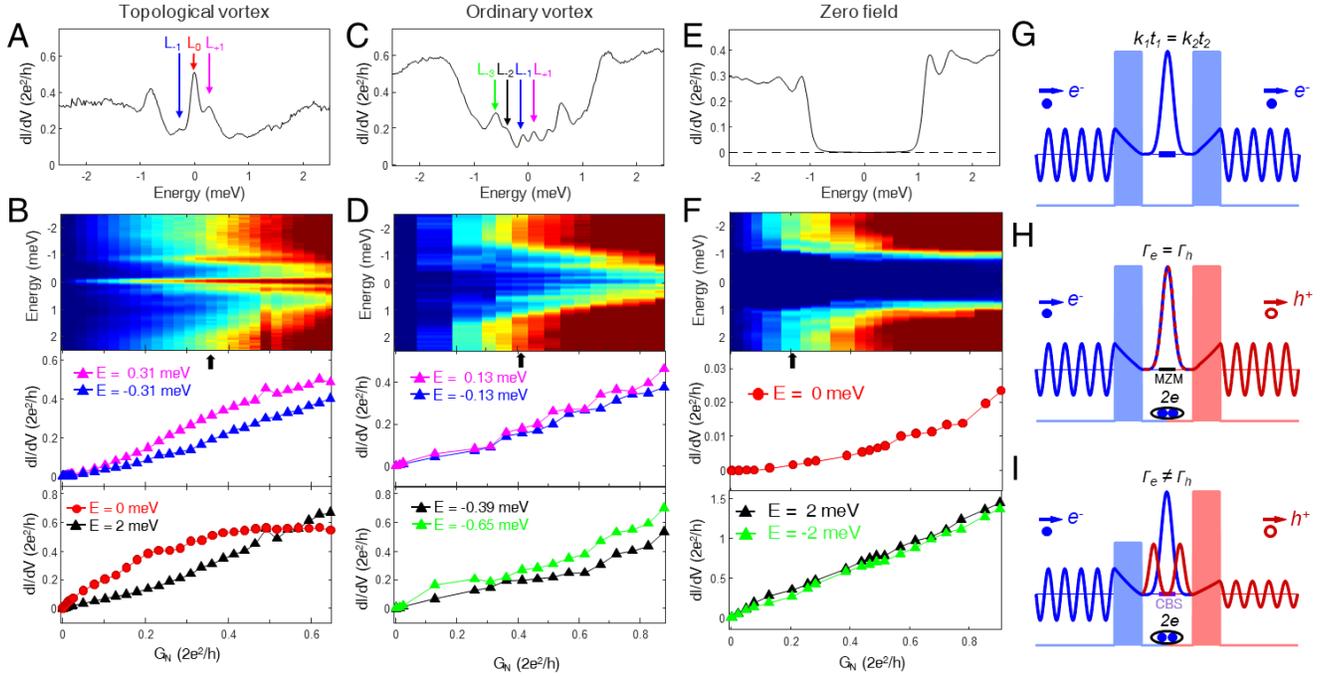

**Fig. 2. Majorana induced resonant Andreev reflection.** (**A**) A *dI/dV* spectrum measured at the center of a topological vortex ($V_s$ = -5 mV, $I_t$ = 140 nA, $V_{mod}$ = 0.02 mV), which shows an MZM coexisting with CBSs located at ±0.31 meV (marked by mauve and blue arrows). (**B**) A tunnel coupling dependent measurement on the vortex shown in (A) at 2 T. Top panel: a color-scale plot, *dI/dV(E, $G_N$)*. The $G_N$ position of (A) is marked by a black arrow. Middle panel: tunnel coupling evolution of CBS conductance, which shows no plateau behavior. Bottom panel: tunnel coupling evolution of conductance at the energies of 0 meV (red circles) and 2 meV (black triangles). The observed zero-bias conductance plateau and the increasing conductance at the high bias values (Fig. 1) are reproduced in this vortex core, with a zero-bias $G_P$ equals to 0.54 $G_0$. (**C**) A *dI/dV* spectrum measured at the center of an ordinary vortex ($V_s$ = -5 mV, $I_t$ = 160 nA, $V_{mod}$ = 0.02 mV), which clearly shows three levels of CBS at ±0.13 meV (mauve and blue arrows), ±0.39 meV (black arrow) and ±0.65 meV (green arrow). (**D**) A tunnel coupling dependent measurement on the vortex shown in (C). Top panel: a color-scale plot, *dI/dV(E, $G_N$)*. The $G_N$ position of (C) is marked by a black arrow. Middle panel and bottom panel: tunnel coupling evolution of CBS conductance. It is confirmed that there is no plateau feature for the CBS in the ordinary vortex. (**E**) A *dI/dV* spectrum measured at 0 T ($V_s$ = -5 mV, $I_t$ = 80 nA, $V_{mod}$ = 0.02 mV). A hard superconducting gap can be seen. (**F**) Top panel: a color-scale plot of tunnel coupling dependent measurement of (E). The $G_N$ position of (E) is marked by a black arrow. Middle panel: tunnel coupling evolution of zero-bias conductance (normal metal - superconductor junction case). Bottom panel: tunnel coupling evolution of conductance at the above gap energy (normal metal - normal metal junction case). There is no plateau behavior on 0 T which indicates that our measurements are free of quantum ballistic transport. (**G**) A schematic plot of resonant tunneling through a symmetric double barrier system. The wavefunction evolution of a tunneled electron is shown. *kt* is penetration constant, which is defined as $kt = t\sqrt{2m(V - E_q)}/h$, *t* being the barrier width, *V* the barrier potential, and $E_q$ the energy of the quasistationary state. At $E_q$, the reflected electrons from the inside of double barrier interfere destructively with the incident electrons, realizing a perfect transmission regardless of strength of tunnel barriers as long as the same *kt* for the two barriers. (**H**) The double-barrier view of the MZM-induced resonant Andreev reflection. The blue and red colors indicate the electron and hole process, respectively.

An MZM has equivalence of particle and hole components which ensures the same tunnel coupling on electron and hole barrier ($\Gamma_e = \Gamma_h$). Thus, as an analogue of (E), a resonant Andreev reflection will happen on the particle-hole Hilbert space mediated by the MZM. **(I)** The double-barrier view of Andreev reflection mediated by a CBS. The arbitrary mixing of particle-hole components in CBS breaks the resonance condition.

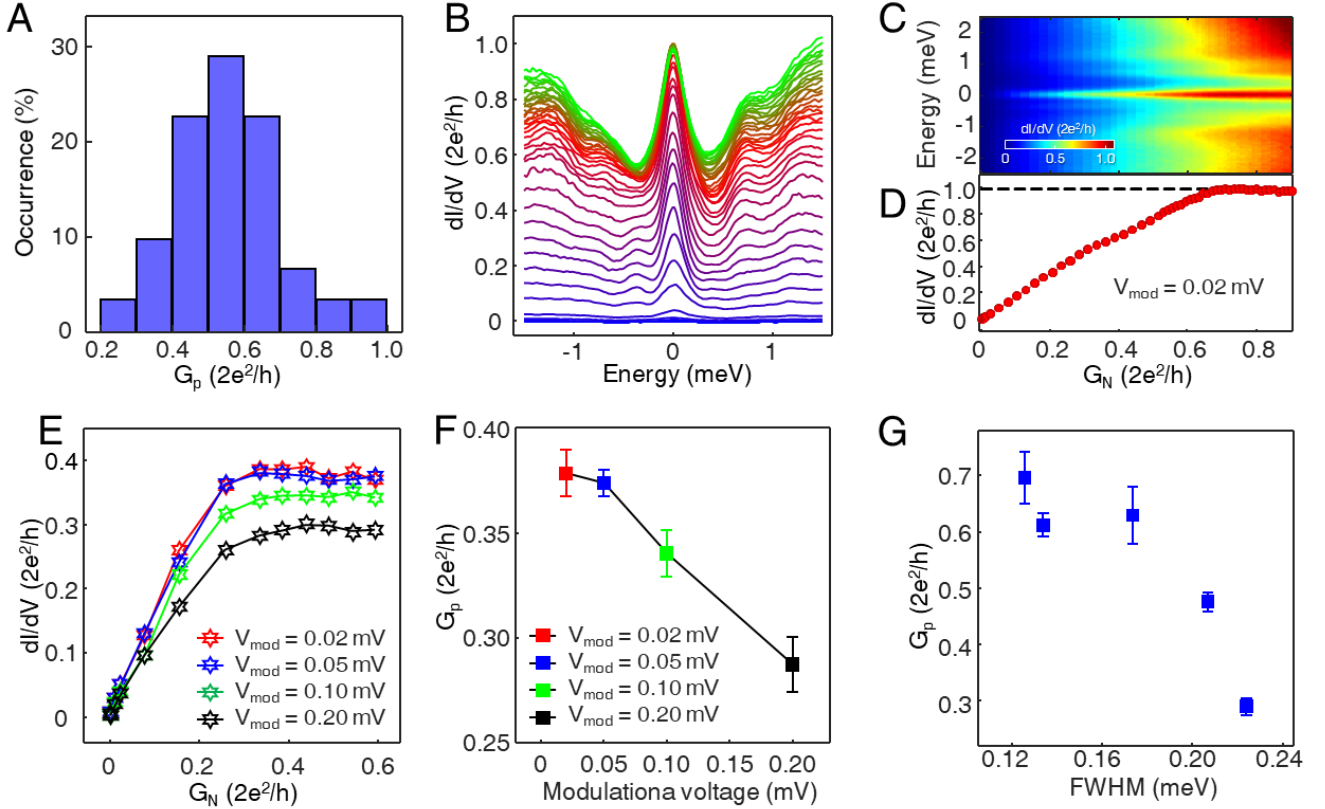

**Fig. 3. The conductance variation of Majorana plateau. (A)** A histogram of the $G_p$ from 31 sets of data which are measured with the same instrument. Sorting of the plateau conductance ($G_p$) in the order of increasing magnitude can be found in Fig. S9 in (*26*). ($V_s$ = -5 mV, $V_{mod}$ = 0.02 mV) **(B)** The overlapping plot of 38 $dI/dV$ spectra selected from a topological vortex which reaches quantized conductance plateau. ($V_s$ = -5 mV, $V_{mod}$ = 0.02 mV) **(C)** A color-scale plot of (B) within the energy range of [-2.5, 2.5] meV that expands the spectra as a function of $G_N$. **(D)** A horizontal line-cut at the zero-bias from (C). The conductance curve shows a conductance plateau achieved $G_0$. **(E)** A series of tunnel coupling dependent measurements on the same MZM, with four modulation voltages of 0.02 mV, 0.05 mV, 0.10 mV and 0.20 mV. **(F)** The plot of $G_P$ as a function of modulation voltage of the data shown in (E). **(G)** Relationship between FWHM of ZBCP and $G_P$, obtained from five different MZMs measured at the same experimental conditions, suggesting that the quasiparticle poisoning effect affect the plateau value. The FWHM were extracted from the spectrum measured at a large tip-sample distance with the same experimental parameters ($V_s$ = -5 mV, $I_t$ = 500 pA, $V_{mod}$ = 0.02 mV).

# Supplementary Materials for

# Nearly quantized conductance plateau of vortex zero mode in an iron-based superconductor

Shiyu Zhu, Lingyuan Kong, Lu Cao, Hui Chen, Michał Papaj, Shixuan Du, Yuqing Xing, Wenyao Liu, Dongfei Wang, Chengmin Shen, Fazhi Yang, John Schneeloch, Ruidan Zhong, Genda Gu, Liang Fu, Yu-Yang Zhang[*], Hong Ding[*], and Hong-Jun Gao[*]

[*]Correspondence to: hjgao@iphy.ac.cn, dingh@iphy.ac.cn, zhangyuyang@ucas.ac.cn

**This PDF file includes:**

- Materials and Methods
- Supplementary Text
- Figs. S1 to S9
- References

**Materials and Methods:**

Large single crystals of FeTe$_{0.55}$Se$_{0.45}$ with high quality were grown using the self-flux method, and their values of T$_c$ were determined to be 14.5 K from magnetization measurements (*45*). There are two kinds of single crystals crystallizing simultaneously with similar structure and Te/Se compositions. Fe$_{1+y}$Te$_{0.55}$Se$_{0.45}$ single crystals with excess Fe atoms, with shining surfaces and being easy to cleave, are non-superconducting before annealing under Te atmosphere. FeTe$_{0.55}$Se$_{0.45}$ single crystals without excess Fe, usually without shining surface, are superconducting without post-annealing. All STM/S data shown in this paper are from as-grown FeTe$_{0.55}$Se$_{0.45}$ single crystals, and our previous ARPES data (*15*) and STM data (*16*) are also from this kind of samples. The samples used in the experiments were cleaved *in situ* and immediately transferred to an STM head.

Experiments were performed in two ultrahigh vacuum (1 × 10$^{-11}$ mbar) LT-STM systems, STM#3 (DR 30 mK, (9-2-2) T) and STM#2 ($^3$He 400 mK, (9-2-2) T). All the STM data are measured with tungsten tips under a setpoint voltage of -5 mV. The tunnel-coupling dependent measurements are measured by fixing the setpoint voltage at -5 mV and varying the tunneling conductance with a close-loop feedback. The z-offset value is recorded by the piezoelectric ceramic during the measurements. Unless otherwise noted, the differential conductance (*dI/dV*) spectra were acquired by a standard lock-in amplifier at a frequency of 973.1 Hz under the modulation voltage, the zero-to-peak amplitude of lock-in excitation, $V_{mod}$ = 0.02 mV for STM#3, and a frequency of 973.0 Hz under the modulation voltage $V_{mod}$ = 0.10 mV for STM#2. The effective electronic temperature for STM #3 is 377 mK as discussed in (I) part below. A vector magnetic field with the maximum value 9$_z$-2$_x$-2$_y$ Tesla can be applied to a sample for both STM#3 and STM#2. Data in Fig. S7 were measured by STM#2, while others were measured by STM#3.

**Supplementary Text**

I. Determination of effective electron temperature and energy resolution.

We determined the effective electron temperature ($T_{eff}$) by measuring the superconducting Al(111) single crystal for STM#3. We fit the dI/dV spectra by the theoretical formula as following:
$$g(V) = -\int_{-\infty}^{\infty}\left[\int_{-\infty}^{\infty}\rho(E)f'(\epsilon+E)dE\right]b(eV-\epsilon)d\epsilon$$
This equation represents the density of states from sample $\rho(E)$ convolved with the thermal broadening f'(E) and bias modulation broadening b(V) (46). Here, we adopt Dynes function $\rho_D(E)$ for the sample (47).
$$\rho_D(E) = \text{Re}\left(\frac{E-i\Gamma}{\sqrt{(E-i\Gamma)^2-\Delta^2}}\right)$$
The f'(E) is the derivative of Fermi function and b(V) is the known lock-in broadening function,
$$f'(\epsilon) = \frac{df(\epsilon)}{d\epsilon} = -\frac{\beta\exp(\beta\epsilon)}{[\exp(\beta\epsilon)+1]^2}$$

$$b(V) = \begin{cases} \dfrac{2}{\pi V_{mod}}\sqrt{1-(\dfrac{V}{V_{mod}})^2}, & (|V| < V_{mod}) \\ 0, & (|V| > V_{mod}) \end{cases}$$

where $\beta = (k_B T_{eff})^{-1}$ and $V_{mod}$ is the zero-to-peak amplitude of lock-in modulation voltage. The spectra on Al were measured using the modulation voltage $V_{mod} = 0.005$ meV.

The superconducting spectrum of Al was shown in Fig. S1. By fitting with the above formulas, we get the effective electron temperature $T_{eff} = (377 \pm 30)$ mK, corresponding to an energy resolution of $(0.114 \pm 0.009)$ meV.

II. Zero-bias calibration

In a tunnel-coupling dependent measurement, the tunneling conductance changes about three orders of magnitude (from $1 \times 10^{-3}$ $G_0$ to 0.9 $G_0$ in the case of Fig. 1). The large tunneling conductance variation brings a recognizable zero-bias shift of the instrument with changing tip-sample distance. Therefore, the zero-bias calibration is needed. We plot the peak energy ($E_p$) as a function of tunneling conductance, as shown in Fig. S2A.

The fluctuation range of peaks is much smaller than the energy resolution, according to the calibration in part (I). Therefore, we calibrate the peaks back to zero-bias by shifting the x-axis without interpolation. The raw data without any processing are shown in Fig. S2B, which are corresponding to the data of Fig. 1 in the main text.

III. The z-offset plot of zero-bias conductance peak

In addition to the plot of ZBCP peak height as a function of $G_N$ in Fig. 1F, we also plot the ZBCP peak height as a function of z-offset in Fig. S3A. In the z-offset plot, the tunneling conductance of ZBCP peak height should exponentially increase at the weak coupling region, which is confirmed by the observation. The z-offset values are converted from the recording voltage of piezoelectric ceramic at z-direction when measuring $dI/dV$ curves. The z-offset-dependent plot displays a plateau feature similar to the one shown in Fig. 1F. Moreover, the conductance of the above-gap state increases as a continuous exponential form without any steps, confirming no ballistic transport occurs in the STS measurement. As a comparison, a z-offset plot measured at 0 T on FeTe$_{0.55}$Se$_{0.45}$ surface is shown in Fig. S3B. Both the zero-bias and the high-bias conductance keep increasing with z-offset, implying a tunneling process without ballistic transport.

IV. Tuning instrumental broadening by varying the modulation voltage

We changed the instrumental broadening ($r$) by varying the lock-in modulation voltage, which contributes as the lock-in voltage broadening ($r_{lock-in}$). The lock-in voltage broadening can be present by $b(V)$ (see part I). We estimate the total broadening by

$$r = \sqrt{r_{temperature}^2 + r_{lock-in}^2}$$

where $r_{temperature}$ is thermal broadening, which is calculated by the effective electron temperature.

The modulation voltage dependent measurements are shown in Figs. 3E-F. The four sets of barrier-dependent data are measured with the zero-to-peak amplitude of lock-in modulation voltages of 0.02 mV, 0.05 mV, 0.10 mV and 0.20 mV, respectively. They have the instrumental broadening of 0.12 meV, 0.12 meV, 0.15 meV and 0.23 meV, respectively. Increasing the modulation voltage leads to a decreasing plateau conductance of ZBCPs.

V. Reproducibility of conductance plateau

As shown in Figs. 2C-D, not all the vortices possess an MZM because of the large inhomogeneity in Fe(Te,Se). Accurately, when we measured FeTe$_{0.55}$Se$_{0.45}$ by STM#3 under 2 T magnetic field, there are about 50% vortices possessing an MZM. Among them, about 50% vortices are stable enough for completion of a tip-approaching process and observing a plateau feature, while others are pushed away during the process. Once a vortex survives during the tip-approaching process, its plateau feature is usually robust and reproducible. Figure S5 shows the detailed data of Fig. 3G. We also plot four more sets of data in Fig. S5. These data are measured under the same experimental parameters by STM #3. We also checked the conductance plateau feature using STM #2. Two sets of data are shown in Fig. S7. It is clear that the plateau feature is also robust.

VI. Extracting the full width of half maximum (FWHM) of MZM

We define the formula of an MZM peak using the $g(V)$ described in part I, by adopting a Lorentzian for the sample density of state $\rho(E)$. The peak is considered as the convolution of a Lorentz function (intrinsic property), a derivative Fermi function (thermal broadening effect), and a semi-circle function (modulation broadening). The full width of half maximum (FWHM) was extracted directly from the fitting curve of the MZM peak.

VII. Reversibility of tunnel coupling dependent measurements.

The STM tips are made to be stable enough before a tunnel coupling dependent measurement that the tip state does not change during the tip-approaching process. That gives us an opportunity to reproduce the plateau feature by a second measurement at the same condition. A set of reversibility data of tunnel coupling dependent measurements is shown in Fig. S8. The zero-bias conductance map and atomic resolved topography measured before (Figs. S8A & S8B) and after (Figs. S8C & S8D) the tip-approaching processes are nearly identical, indicating the non-changed STM tip and the sample surface during the measurement. The zero-energy and high-energy conductance evolution shown in Fig. S8E and Fig. S8F are measured by two successive tip-approaching processes. The two sets of data show the similar average plateau conductance of 0.30 $G_0$, providing an excellent repeatability of the measurements. However, there are also some obstacles to get a perfect repeat, such as the enhancement of vortex creep rate when increasing the tunnel-barrier conductance. The jump of a vortex always induces a discontinuous change of $dI/dV$ curves, which interrupts the measurement but does not damage the STM tip.

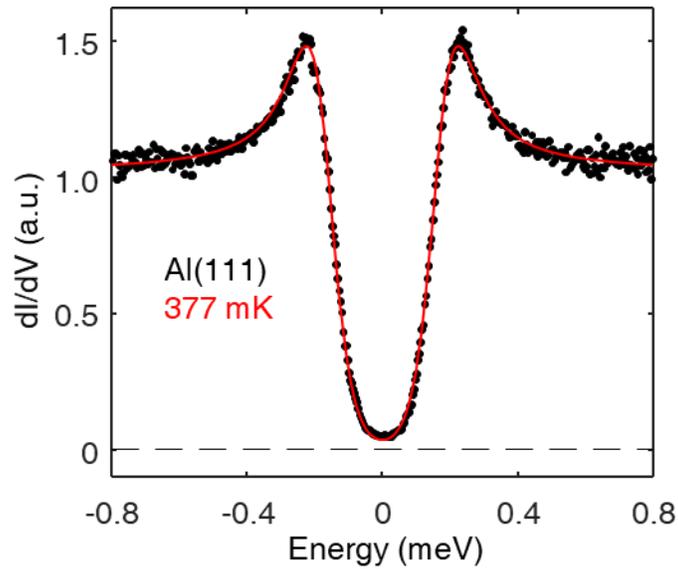

**Fig. S1. Determination of effective electron temperature.** Superconducting spectra measured on Al(111), showing an effective electron temperature of 377 mK.

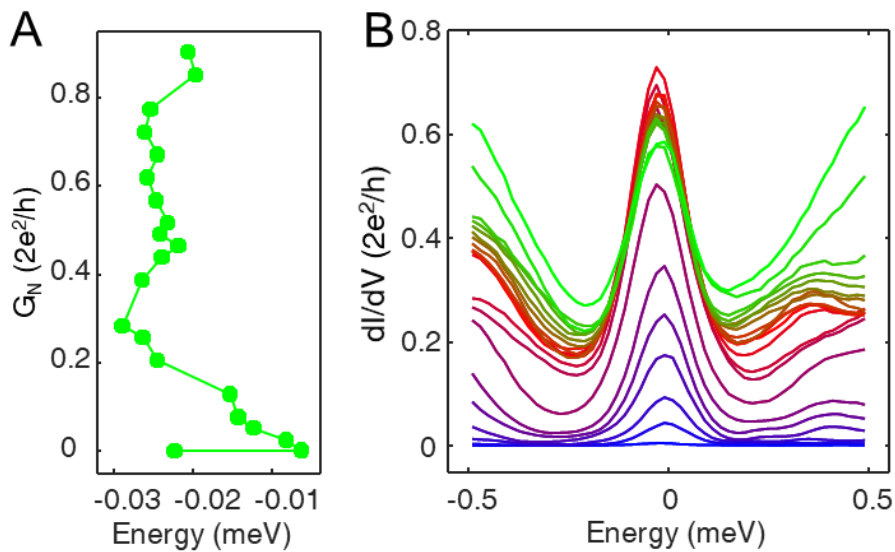

**Fig. S2. Zero-energy calibration. (A)** Peak energies of the data shown in Fig. 1 before zero-bias calibration measured by STM#3. The peak energy fluctuates range is less than 0.03 meV. **(B)** The overlap plot of STS raw data corresponding to that shown in Fig. 1. For clarity, only the data points in the energy range of [-0.5, 0.5] meV are shown.

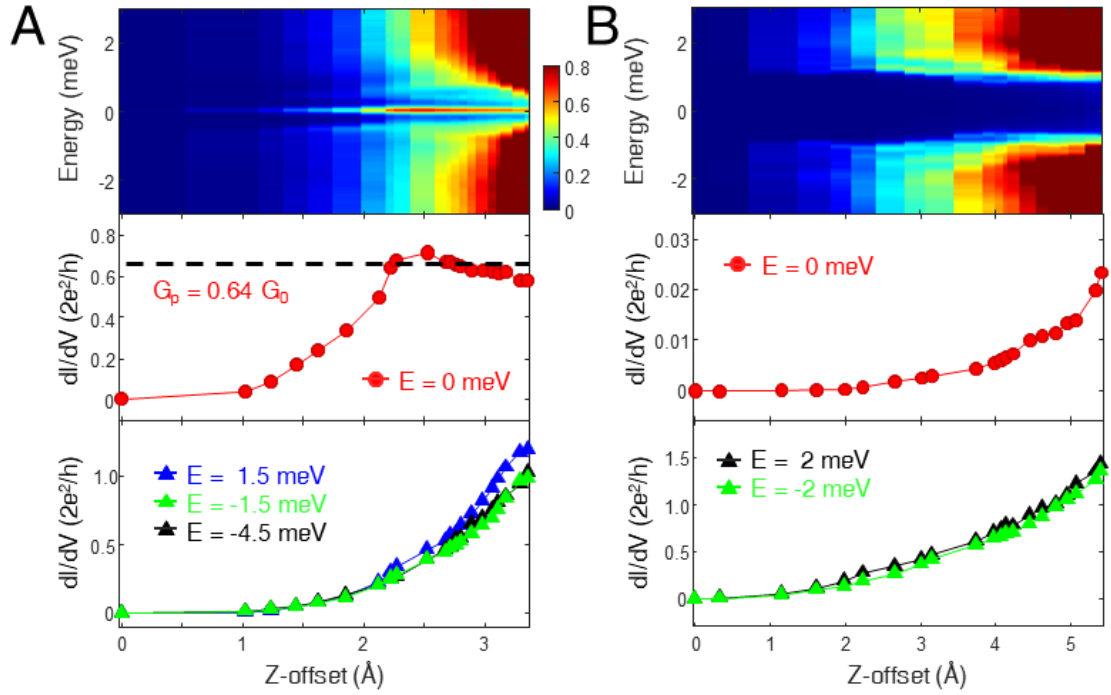

**Fig. S3. Conductance plot as a function of z-offset.** (**A**) Z-offset plots of zero-bias and high-bias conductance behaviors (2 T), which correspond to the data of Fig. 1 in the main text. (**B**) Z-offset plots of zero-bias and high-bias conductance, measured at 0 T on $FeTe_{0.55}Se_{0.45}$. The exponential increasing feature confirms that there is no ballistic process.

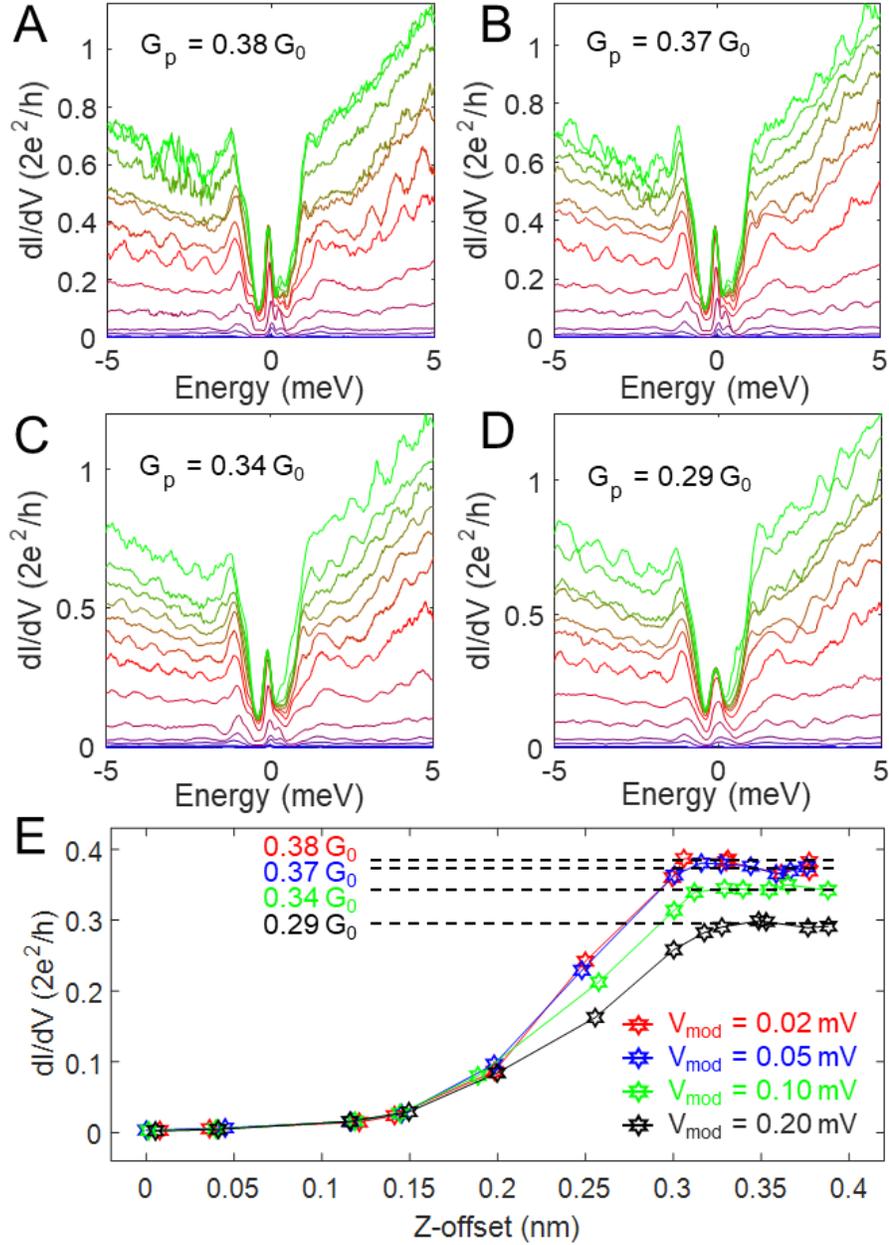

**Fig. S4. Tunnel coupling dependent measurements at the same vortex with different modulation voltages. (A)-(D)** Overlapping plots of STS raw data which are measured with the modulation voltage of 0.02 mV, 0.05 mV, 0.10 mV and 0.20 mV, respectively. The raw data correspond to the processed data of Fig. 3B in the main text. **(E)** Zero-bias conductance plot of (A-D) as a function of z-offset, which shows the variation of plateau value for different modulation voltages.

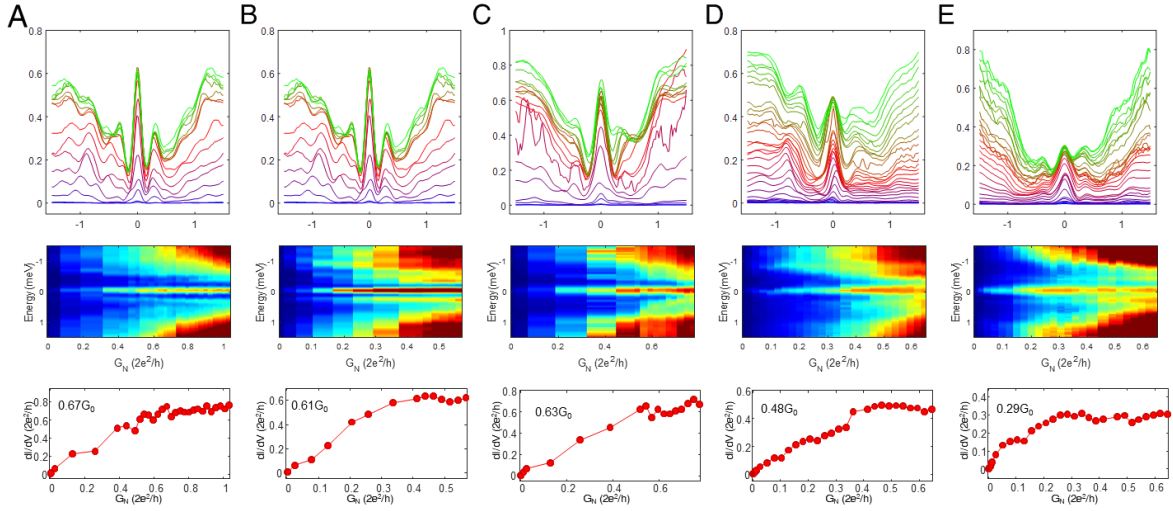

**Fig. S5. Tunnel-coupling dependent measurements on different MZMs.** Overlapping plots, color-scaled plots and zero-bias conductance curves of five MZMs are presented. The five sets of data correspond to the data shown in Fig. 3G.

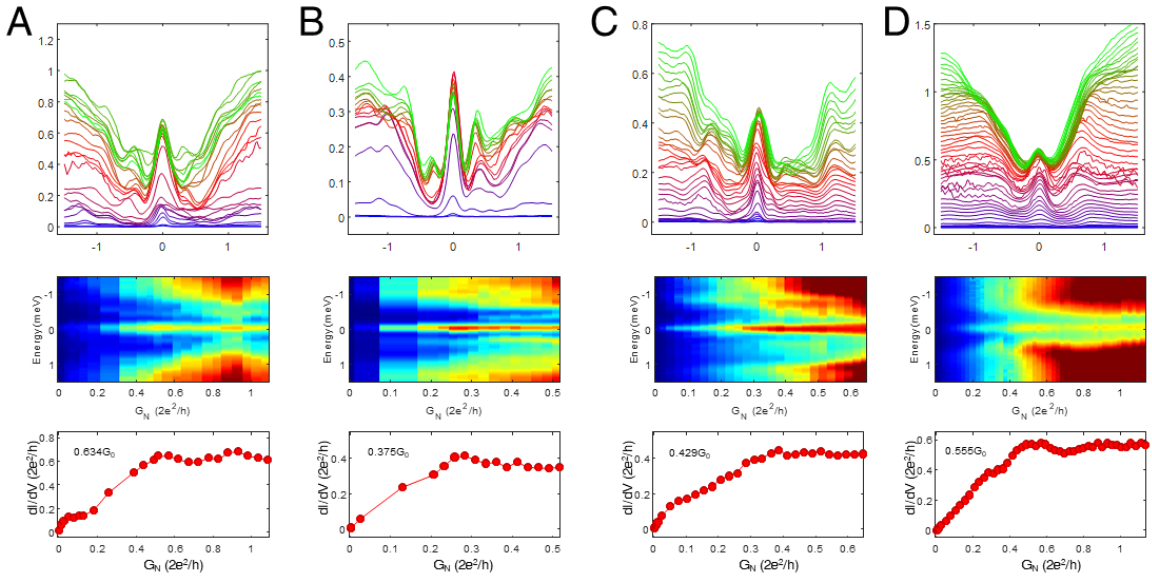

**Fig. S6. Tunnel-coupling dependent measurements at different samples and areas reflecting the reproducibility of the plateau feature.** Overlapping plots, color-scaled plots and zero-bias conductance curves of four more data are presented, which have already been included in Fig. 3A.

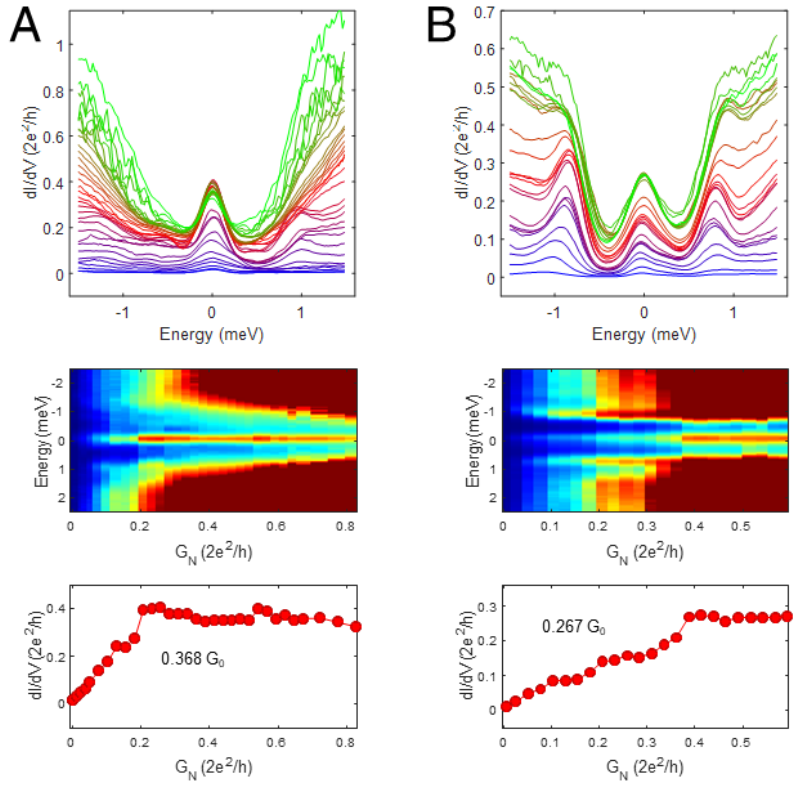

**Fig. S7. Tunnel-coupling dependent measurements on STM #2.** Overlapping plots, color-scaled plot and zero-bias conductance curves of two sets of data are presented, which are measured on STM #2. The effective electron temperature is 906 mK.

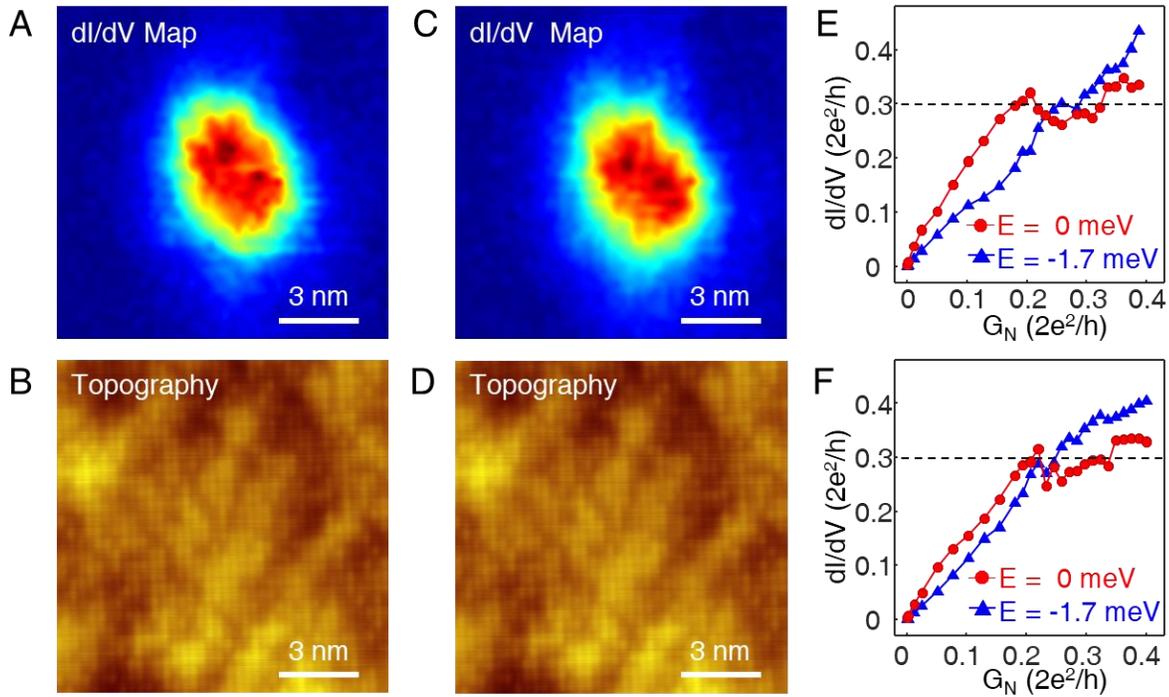

**Fig. S8. Reversibility of tunnel coupling dependent measurements. (A)-(B)** A zero-bias dI/dV map and corresponding STM topography measured before tunnel coupling dependent measurements. The map and the topography are measured at the same area. The magnetic field is 2.0 T. **(C)-(D)** A zero-bias dI/dV map and corresponding STM topography measured after tunnel coupling dependent measurements. The magnetic field is 2.0 T. The measuring parameters are the same with the ones in (A-B): sample bias, $V_s$= –5 mV; tunneling current, $I_t$= 500 pA. **(E)-(F)** Two repeated sequences of tunnel coupling dependent measurements at the same spatial position, showing an average plateau conductance of 0.30 $G_0$, respectively. The data shown in (F) are recorded during a second tip-approaching process after finishing the first one.

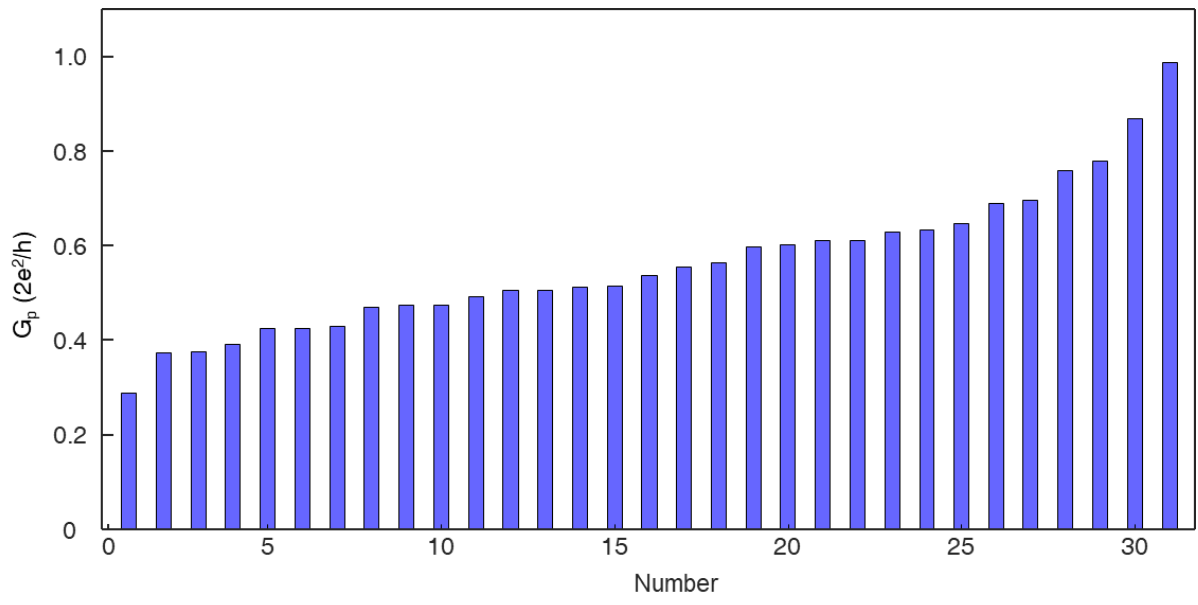

**Fig. S9.** Sorting of the plateau conductance ($G_p$) in the order of increasing magnitude.